\newcommand{\be}{\begin{equation}}\newcommand{\ee}{\end{equation}}
\newcommand{\bea}{\begin{eqnarray}}\newcommand{\eea}{\end{eqnarray}}
\newcommand{\nn}{\nonumber}\newcommand{\p}[1]{(\ref{#1})}
\newcommand{\lb}[1]{\label{#1}}
\newcommand\s{\scriptscriptstyle}

\newcommand\q{\quad}
\newcommand\qq{\quad\quad}

\newcommand\cB{{\cal B}}

\newcommand\cF{{\cal F}}

\newcommand\cP{{\cal P}}

\def\a{\alpha}
\def\da{{\dot\alpha}}
\def\b{\beta}
\def\db{{\dot\beta}}
\def\g{\gamma}
\def\dg{{\dot\gamma}}
\def\d{\delta}

\def\eps{\epsilon}
\def\bep{{\bar\epsilon}}          
\def\ve{\varepsilon}

\def\bph{{\bar\phi}}

\def\l{\lambda}
 \def\th{\theta}  \def\bt{\bar\theta}
\def\vt{\vartheta}
\def\r{\rho}

\def\si{\sigma}  
\def\bs{\bar\sigma}

\def\D{\Delta}

\def\pa{\partial}

\newcommand\ada{{\alpha\dot{\alpha}}}

\newcommand\bdb{{\beta\dot{\beta}}}
\newcommand\bda{{\beta\dot{\alpha}}}

\newcommand\ab{{\alpha\beta}}

\newcommand\pada{\partial_{\alpha\dot{\alpha}}}

\newcommand\R{{\s R}}
\newcommand\M{{\s M}}
\newcommand\sL{{\s L}}
\newcommand\N{{\s N}}

\newcommand\hd{{\hat\delta}}

\newcommand\hL{{\hat{L}}}

\newcommand\hO{{\hat{O}}}

\newcommand\hX{{\hat{X}}}
\newcommand{\pp}{{\s ++}}

\newcommand{\Dp}{D^{\pp}}

\newcommand{\Vp}{V^\pp}

\def\sfrac#1#2{{\textstyle\frac{#1}{#2}}}

\documentclass[12pt]{article}
\usepackage{amsmath,amssymb}

\renewcommand{\thefootnote}{\fnsymbol{footnote}}

\begin{document}
\begin{center}
{\bf  TWIST-DEFORMED SUPERSYMMETRIES IN\\ NON-ANTICOMMUTATIVE SUPERSPACES}\\
\vspace{1cm}

{\large\bf
  B.M. Zupnik\footnote{zupnik@theor.jinr.ru}}
\vspace{1cm}

{\it Bogoliubov Laboratory of Theoretical Physics, JINR, \\
141980, Dubna, Moscow Region, Russia}\\
\end{center}

\begin{abstract}
We consider a quantum group interpretation of the non-anticommutative deformations
in Euclidean supersymmetric theories. Twist deformations in the corresponding 
superspaces and  supergroups are constructed in terms of the left supersymmetry 
generators. Non-anticommutative $\star$-products of superfields are covariant 
objects in the twist-deformed  supersymmetries, and this covariance guarantees 
the manifest invariance of superfield actions using $\star$-products.
\end{abstract}


\renewcommand{\thefootnote}{\arabic{footnote}}
\setcounter{footnote}0
\setcounter{equation}0
\section{Introduction }

The most popular classical noncommutative field theory (see, e.g. review 
\cite{Rev1}) can be realized on ordinary smooth field functions $f(x), g(x)$ 
on $R^4$ using the following pseudolocal representation of the $\star$-product :
\bea
&&f\star g=f e^P g=fg+\sfrac{i}2\vt_{mn}\pa_mf\pa_ng-\sfrac18\vt_{mn}\vt_{rs}\pa_m
\pa_r f\pa_n\pa_sg+\ldots,\nn\\
&&\lb{star} fPg=\sfrac{i}2\vt_{mn}\pa_mf\pa_ng.
\eea
where $x_m$ are the coordinates of $R^4$, $\pa_m=\pa/\pa x_m$, and $\vt_{mn}$ are 
some constants ( $m,n=1,2,3,4$). All products of the functions and their 
derivatives in the right-hand side are commutative.  It is evident that nonlinear 
interactions in these noncommutative (nonlocal) field theories are not invariant 
with respect to the standard Lorentz transformations of local fields.
 
The quantum group structures in this noncommutative algebra of functions were found
and analyzed in  \cite{Oe}-\cite{KoM}. The basic point of this interpretation
is connected with the twist operator acting on tensor products of functions
\be
\cF =\exp(\cP),\q \cP=\sfrac{i}{2}\vt^{mn}P_m\otimes P_n\lb{twist}
\ee
where $P_m f=\pa_m f$.
The strict definition of the noncommutative product is 
\be
f\star g=\mu\circ \cF f\otimes g,\q \mu\circ f\otimes g=fg\lb{startw}
\ee
where $\mu$ is the multiplication map in the commutative algebra. Thus, this  
twist operator is the quantum-group analog of the pseudolocal operator $\exp(P)$ 
\p{star}.

Let us consider  generators of the Poincar\'e group $P_m$ and $M_{mn}$. By 
definition, the twist-deformed  Poincar\'e  group $U_t(P_m,M_{mn})$ has the 
undeformed Lie algebra of generators; however, its coproduct is deformed 
\bea
&&\D_t(P_m)=P_m\otimes 1+1\otimes P_m,\nn\\
&&\D_t(M_{mn})=\exp(-\cP)(M_{mn}\otimes 1+1\otimes M_{mn})\exp(\cP).\lb{coprM}
\eea 

The exact constructions of maps between differential operators on commutative 
and noncommutative algebras of functions were formulated in recent papers of 
the Munchen group \cite{We}. It was shown that the $\star$-product \p{startw} 
transforms covariantly in $U_t(P_m,M_{mn})$, but the Leibniz rule for  deformed 
transformations is changed according to Eq.\p{coprM}.  4D-space integrals of 
the covariant $\star$-products of fields are invariant with respect to 
$U_t(P_m,M_{mn})$. The quadratic free interactions possess also the standard 
Poincar\'e invariance.

We shall consider the quantum group interpretation of the non-anticommutative
deformations in the Euclidean supersymmetric theories \cite{Se}-\cite{FS} 
\footnote{Note that the alternative quantum-group deformations of  supersymmetries 
with a more complex supersymmetric geometry were considered earlier \cite{KMLS}, 
however, we shall not discuss these models here.}. The basic $\star$-product of 
these models is realized on the standard local superfields, and supersymmetry 
generators can be presented as the 1-st order differential operators on the 
undeformed superspace. The left-handed Grassmann coordinates  of these superspaces 
do not anticommute with respect to the $\star$-product, but the basic chiral 
bosonic coordinates commute with all superspace coordinates. The 
non-anticommutative superspace  is defined exactly as the $\star$-product 
algebra on ordinary functions of the superspace coordinates. The twist 
elements for the nilpotent deformations can be constructed in terms of the left 
supersymmetry generators. 

Section 2 is devoted to the analysis of the twist deformation of the 
$N{=}(\sfrac12,\sfrac12)$ supersymmetry \cite{KS}.  We derive the unusual Leibniz 
rules for the deformed transformations on the products of superfields or the 
products of component fields. Twist deformation of the Euclidean 
$N{=}(1,1)$ supersymmetry in the chiral and harmonic superspaces is considered 
in Sect. 3. The corresponding nilpotent operator $\cP$ is analogous to the basic 
bi-differential operator of the $N{=}(1,1)$ deformations in Refs.\cite{ILZ,FS}. 

The twist interpretation allow us to understand correctly transformation 
properties of $\star$-products of superfields by analogy with the deformed 
transformations of the $\star$-products of fields \p{star} in the noncommutative 
field theory \cite{We}. At the level of the pseudolocal superfield formalism, the  
$t$-supersymmetry is equivalent to the {\it $\star$-covariance principle} for 
the noncommutative algebra of superfields which means the similarity of 
transformations of  local superfields and their $\star$-products. The covariance 
principle allow us to obtain a simplified field-theoretical derivation of the 
unusual Leibniz rules for the deformed supersymmetry transformations on 
$\star$-products of superfields. Known deformed supersymmetric actions in the 
non-anticommutative superspaces are manifestly invariant with respect to the 
corresponding twist-deformed supersymmetry, and this invariance explains naturally 
all selection rules of these theories which could seem formal earlier. 

\setcounter{equation}0
\section{Twist-deformed  $N{=}(\sfrac12,\sfrac12)$ supersymmetry }
We use the  chiral coordinates $z^\M=(y_m, \theta^\a,\bt^\da)$ in the Euclidean 
superspace R$(4|2,2)$, where   $m=1,2,3,4,\q\a=1, 2,$ and $\da=\dot{1},\dot{2}$. 
The central and antichiral 4D coordinates are, respectively,
\be
x_m=y_m-i\th\si_m\bt,\q \bar{y}_m=y_m-2i\th\si_m\bt
\ee
and $(\si_m)_\ada$ are the SO(4) Weyl matrices. Note that these coordinates are 
pseudoreal with respect to the special conjugation \cite{ILZ}
\be
 (y_m)^*=y_m,\q (\th^\a)^*=\ve_\ab \th^\b,\q(\bt^\da)^*=\ve_{\da\db}\bt^\db,
\ee
so one can use the reality condition for the even Euclidean chiral superfield 
$\phi(y,\th)$. The  generators  of the Euclidean  $N{=}(\sfrac12,\sfrac12)$  
supersymmetry SUSY$(\sfrac12,\sfrac12)$ have the following form: 
\bea
&& L^\b_\a=L^\b_\a(y)+L^\b_\a(\th)=\sfrac14(\si_m\bs_n)^\b_\a (y_n\pa_m-y_m\pa_n)
+\th^\b\pa_\a-\sfrac12\d^\b_\a\th^\g\pa_\g,\nn\\
&& R^\db_\da=R^\db_\da(y)+R^\db_\da(\bt)=\sfrac14
(\bs_m\si_n)^\db_\da (y_m\pa_n-y_n\pa_m)+\bt^\db\bar\pa_\da-\sfrac12\d^\db_\da
\bt^\dg\bar\pa_\dg,\nn\\
&&O=\th^\a\pa_\a-\bt^\da\bar\pa_\da,\q Q_\a=\pa_\a,\q \bar Q_\da=\bar\pa_\da
-2i\th^\a\pa_\ada,\q P_m=\pa_m,\lb{difgen}
\eea
where   $(\bs_m)^{\da\a}=\ve^\ab\ve^{\da\db}(\si_m)_\bdb$, and $\pa_\M=(\pa_m, 
\pa_\a,\bar\pa_\da)$ are partial derivatives in the chiral coordinates.
Generators $L^\b_\a, R^\db_\da$ and $O$ correspond to the automorphism group
SU(2)$_\sL\times$SU(2)$_\R\times$O(1,1). The  SUSY$(\sfrac12,\sfrac12)$
transformations  can be separated as follows
\bea
&&\d A=-(g+G)A,\q g=P_c+R_\r+Q_\eps,\q G=L_\l+aO+\bar Q_\bep,\lb{standtr}\\
&&P_c=c_mP_m,\q L_\l=\l^\a_\b L^\b_\a,\q R_\r=\r^\da_\db R^\db_\da,\q 
Q_\eps=\eps^\a Q_\a,\q\bar Q_\bep=\bep^\da\bar Q_\da,\nn
\eea
where the corresponding combinations of operators and transformation parameters 
are introduced. These definitions will be convenient in the deformed supersymmetry.

We shall use the notation S$(4|2,2)$ or C$(4|2,0)$ 
for the supercommutative algebras of general or chiral superfields.
The bilinear multiplication map $\mu$ connects the tensor product of superfields 
with the local supercommutative product in S$(4|2,2)$
\be
\mu\circ A\otimes B=AB=(-1)^{p(A)p(B)}BA.
\ee
The standard coproduct map  is defined on the generators
of SUSY$(\sfrac12,\sfrac12)$  \p{standtr}
$$\D(g)=g\otimes 1+1\otimes g,\qq \D(G)=G\otimes 1+1\otimes G.$$ 
It determines the action of these generators on the tensor 
product of superfields and yields the standard Leibniz rule for supersymmetry 
transformations on the local product of superfields $\d(AB)=(\d A)B+A\d B$.

The non-anticommutative deformation $\hat z=(y_m,\hat\th^\a,\bt^\da)$ of the 
coordinates of the Euclidean $N{=}(\sfrac12,\sfrac12)$ superspace was  
considered in \cite{Se}. The basic operator relation of the non-anticommutative
superspace is
\be
T^\ab(\hat\th)=\hat\th^\a\star\hat\th^\b+\hat\th^\b\star\hat\th^\a-C^\ab=0,
\lb{basicth}
\ee
where $C^\ab$ are some constants. The operator superfields $\hat{A}(y,\hat\th,\bt)$ 
and $\hat{B}(y,\hat\th,\bt)$ with the antisymmetric ordering of the $\hat\th^\a$ 
decomposition contain the highest terms $\sim\ve_\ab\hat\th^\a\star\hat\th^\b$. 
In the pseudolocal representation, we consider the usual superfields $A(z)$ and 
$B(z)$ as the supercommutative images of these operator superfields. 
The corresponding $\star$-product of  superfields $A(z)$ and $B(z)$ is defined via 
the generators  of the left $N{=}(\sfrac12,0)$ supersymmetry
\bea
&A\star B=Ae^P B=A B-\sfrac12(-1)^{p(A)}C^\ab Q_\a A\, Q_\b B-\sfrac{1}{32}
C^\ab C_\ab Q^2A\,Q^2B,&\nn\\
&APB=-\sfrac12(-1)^{p(A)}C^\ab Q_\a A\, Q_\b B,\q P^3=0,\lb{starpr}&
\eea 
where  $p(A)$ is the $Z_2$ grading, and $P$ is the nilpotent 
bi-differential operator.  The deformed algebras S$_\star(4|2,2)$ 
and C$_\star(4|2,0)$ use this noncommutative product for general or chiral 
superfields, respectively. 

The  twist operator in this supersymmetry was  introduced in \cite{KS} (see also
 discussions in \cite{Ku,ISa})
\bea
&&\cF=\exp(\cP),\q \cP=-\sfrac12C^\ab Q_\a \otimes Q_\b,\nn\\
&&\cF(A\otimes B)=A\otimes B -\sfrac12(-1)^{p(A)}C^\ab Q_\a A 
\otimes Q_\b B-\sfrac{1}{32}C^\ab C_\ab Q^2A\otimes Q^2B.
\eea
The bilinear  map $\mu_\star$ in S$_\star(4|2,2)$ can be defined via this twist 
operator
\bea
&&A\star B\equiv\mu_\star\circ A\otimes B=\mu\circ\exp(\cP) A\otimes B,
\eea
so  $\cF$ is the quantum group analog of the pseudolocal operator $e^P$
\p{starpr}.

By  analogy with the map between differential operators on commutative and 
noncommutative algebras of functions \cite{We}, one can easily define the 
corresponding differential operator $\hX_D$ on S$_\star(4|2,2)$ for any 
differential operator $D$ on the supercommutative algebra. In the case of the 
1-st order operator $D_1=\xi^\M(z)\pa_\M$, the image $\hX_{D_1}$ contains, in 
general, terms with  higher derivatives on  S$_\star(4|2,2)$
\bea
&&(\hat\pa_\M\star A)=\pa_\M A,\q (\hat\pa_\M\star z^\N)=\d^\N_\M,\nn\\
&&(\hX_{D_1}\star A)=(D_1 A)=\mu_\star\circ\exp(-\cP)
(\xi^\M(z)\otimes\pa_\M)(1\otimes A)\nn\\
&&=\xi^\M(z)\star\pa_\M A+\sfrac12(-1)^{p(D_1)}C^\ab Q_\a\xi^\M(z)\star\pa_\M 
Q_\b A\nn\\
&&-\sfrac{1}{32}C^\ab C_\ab Q^2\xi^\M(z)\star\pa_\M Q^2 A
\lb{mapdif}
\eea 
where $p(D_1)$ is the $Z_2$ grading of $D_1$.
For instance, the deformed images of generators $\bar Q_\da$ and $L^\b_a$ 
\p{difgen} are the following second-order differential operators on 
S$_\star(4|2,2)$:
\bea
&&(\widehat{\bar{Q}}_\da\star A)=(\bar\pa_\da-2i\th^\a\pada+iC^\ab\pada Q_\b)
\star A=\bar Q_\da A,\nn
\\
&&(\hL^\b_\a\star A)=(L^\b_\a-\sfrac12C^{\b\g}Q_\g Q_\a)\star A=L^\b_\a A,
\lb{hatL}
\eea
while $\hat g=g$ and $\hO=O$. The deformed operators can be used 
in the operator representation of S$_\star(4|2,2)$ to check directly the 
covariance of the basic operator $T^\ab(\hat\th)$ \p{basicth} with respect to 
the transformations of the deformed supersymmetry
 \bea
&&
\widehat{\bar{Q}}_\da\star T^\ab=0,\q\hL^\r_\si\star T^\ab=\d^\a_\si 
T^{\r\b}+\d^\b_\si T^{\a\r}-\d^\r_\si T^{\a\b}.
\eea 

The Lie superalgebra of the deformed generators with hats is isomorphic to
the Lie superalgebra of the undeformed   supersymmetry generators \p{difgen}.

The coproduct $\D_t(G)=e^{-\cP}\D(G)e^\cP$ in SUSY$_t(\sfrac12,\sfrac12)$ is 
deformed on $G=\bar{Q}_\bep+L_\l+aO$, in particular,
\bea
&\D_t(\bar{Q}_\bep)
=\bar{Q}_\bep\otimes 1+1\otimes \bar{Q}_\bep+i\bep^\da C^\ab(\pa_\ada\otimes Q_\b
-Q_\a\otimes \pa_\bda),&\nn\\
&\D_t(L_\l)=L_\l\otimes 1+1\otimes L_\l
+\sfrac12C^{\r\si}(\l^\a_\r Q_\a\otimes Q_\si+\l^\a_\si Q_\r\otimes Q_\a),&\nn\\
&\D_t(O)=O\otimes 1+1\otimes O-C^\ab Q_\a\otimes Q_\b,&\lb{coprO}
\eea
while $e^{-\cP}\D(g)e^\cP=\D(g)=g\otimes 1+1\otimes g$.

Acting by the composition of $\mu_\star$ and coproduct $\D_t(\bar{Q}_\bep)$ on 
the tensor product of superfields, one can obtain  the following relation:
\bea
&&\hd_\bep\star(A\star B)\equiv-\mu_\star\circ\D_t(\bar{Q}_\bep)A\otimes B=
-(\bar{Q}_\bep A)\star B-A\star\bar{Q}_\bep B\nn\\
&&-i\bep^\da C^\ab[(-1)^{p(A)}\pa_\ada A\star Q_\b B-Q_\a A\star \pa_\bda B]=
-\bar Q_\bep (A\star B)\lb{defleib}.
\eea
The last formula can be derived from the pseudolocal definition 
of $A\star B$ \p{starpr}
\be
\hd_\bep\star(A\star B)=-[\bar Q_\bep, (Ae^PB)]=(\bar Q_\bep A)e^PB
+A[\bar Q_\bep,e^P]B+Ae^P(\bar Q_\bep B).
\ee
Note that Eq.\p{defleib} can be treated as the deformed Leibniz rule for 
$\hd_\bep$.

 It is important to formulate the principle of covariance of the algebra 
 S$_\star(4|2,2)$ (or C$_\star(4|2,0)$) with respect to all deformed 
transformations of SUSY$_t(\sfrac12,\sfrac12)$: the primary superfields $A,~B$ 
and their $\star$-product $(A\star B)$ transform analogously
\bea
&&\hat\d_G\star A=
-GA=-\hat G\star A,\nn\\
&& \hat\d_G\star(A\star B)=-G(A\star B)=-\hat G\star(A\star B).
\lb{cov11}
\eea
where $\hat\d_G=(\hd_\bep+\hd_\l+\hd_a)$. In the pseudolocal formalism of 
the noncommutative superfield theory, these relations are 
sufficient to derive the deformed Leibniz rules.

It is evident that the algebra S$_\star(4|2,2)$ is  not covariant with 
respect to the undeformed supersymmetry, e.g., 
$\d_\bep (A\star B) =(\d_\bep A\star B)+(A\star\d_\bep B)\neq 
-\bep^\da\bar Q_\da(A\star B).$

The deformed transformations act  noncovariantly on the supercommutative product 
of superfields $AB$. For instance, it is not difficult to consider the following 
transformations of the ordinary  product of the even chiral superfields: 
\bea
&&\hd_\bep\star(\phi_1 \phi_2)\equiv -\mu\circ\D_t(\bar{Q}_\bep)\phi_1\otimes
\phi_2 = -\bar Q_\bep (\phi_1\phi_2)-i\bep^\da C^\ab(\pa_\ada \phi_1\, 
Q_\b \phi_2\nn\\
&&-Q_\a \phi_1\, \pa_\bda \phi_2)=\hd_\bep(a_1a_2)+O(\th),\nn\\
&&\hd_\l\star(\phi_1 \phi_2)\equiv -\mu\circ\D_t(L_\l)\phi_1\otimes\phi_2=
-\l^\a_\b L^\b_\a(\phi_1 \phi_2)\nn\\
&&-\sfrac12 C^{\r\si}(\l^\a_\r Q_\a\phi_1 Q_\si\phi_2+\l^\a_\si Q_\r\phi_1 
Q_\a\phi_2)=\hd_\l(a_1a_2)+O(\th).\lb{noncov1}
\eea
The first terms  coincide with the transformations of the undeformed supersymmetry.
Using the $\th$-decomposition of these superfield formulas one can obtain the  
deformed transformations of the products of component fields, for instance,
\bea
&& \hd_\bep(a_1a_2)=-i\bep^\da C^\ab(\pada a_1\,\psi_{\b 2}-\psi_{\a 1}
\pa_{\b\da}a_2),\nn\\
&&\hd_\l(a_1a_2)=-\l^\a_\b L^\b_\a(y)(a_1a_2)
-\sfrac12C^{\r\si}(\l^\a_\r \psi_{\a 1} \psi_{\si 2}
+\l^\a_\si \psi_{\r 1} \psi_{\a 2}) .\lb{noncov2}
\eea

Let us consider two even chiral superfields $\phi_1$ and $\phi_2$ in the chiral 
basis
\bea
&&\phi_i=a_i+\th^\a\psi_{i\a}+\th^2f_i,\\
&&Q_\a \phi_i=\psi_{i\a}+2\th_\a f_i,\q Q^2\phi_i=-4f_i.\nn
\eea
The $\th$-decomposition of the $\star$-product of two chiral superfields depends on
these components and constants $C^\ab~$
\bea
&&\Phi_{12}=\phi_1\star\phi_2=\cB+\th^\a\Psi_\a +\th^2F,\lb{defcomp}
\\
&&\cB=a_1a_2-\sfrac12C^\ab\psi_{1\a}\psi_{2\b}-\sfrac12C^\ab C_\ab f_1f_2,\nn\\
&&\Psi_\a=a_1\psi_{2\a}+a_2\psi_{1\a}-C_\ab(f_1\psi_2^\b-f_2\psi_1^\b),\nn\\
&&F=a_1f_2+a_2f_1-\sfrac12\psi^\a_1\psi_{2\a}.
\eea
These relations can be treated as a deformed tensor calculus for chiral component
multiplets. The $t$-supersymmetry transformations of the composite components 
\p{defcomp} are completely analogous to the  
transformations of the basic components $a_i, \psi_{\a i}, f_i$
\bea
&&\hd_\bep \cB=0,\q \hd_\bep\Psi_\a=-2i\bep^\da\pada\cB,\q
\hd_\bep F=-i\bep^\da\pada\Psi^\a,\nn\\
&&\hd_\l\cB=-\l^\a_\b L^\b_\a(y)\cB,\q \hd_\l\Psi_\g=\l^\a_\g\Psi_\a-
\l^\a_\b L^\b_\a(y)\Psi_\g,\q
\hd_\l F=-\l^\a_\b L^\b_\a(y)F .\lb{compcov}
\eea
These transformations are compatible with the noncovariant component 
transformations \p{noncov2}.

 The non-anticommutative deformation of the Euclidean model for an arbitrary 
number of the  chiral and antichiral superfields $\phi_a$ and  $\bph_a$ 
is based on the superfield action $S_\star(\phi_a,\bph_a)$ \cite{Se}. 
Each term of the $\star$-polynomial decomposition of this  action 
 is separately invariant with respect to  SUSY$_t(\sfrac12,\sfrac12)$, while 
 the quadratic terms  like $\int d^8z\, \phi_a\star
\bph_a=\int d^8z\,\phi_a\bph_a$ possess also the undeformed supersymmetry.

\setcounter{equation}0
\section{Twist-deformed  $N{=}(1,1)$ supersymmetry } 

The nilpotent deformations of the Euclidean $N{=}(1,1)$ supersymmetry were 
considered in the framework of the harmonic-superspace  approach \cite{ILZ,FS}. 
Harmonic-superspace coordinates contain  the  SU(2)/U(1) harmonics $u^\pm_i$ 
and the  chiral superspace coordinates
\be
z^\M=(y_m, \th^\a_k, \bt^{\da k})\, , \q y_m = x_m+
i\th_k\si_m\bt^k
\ee
where $x_m$ are the central 4D coordinates.
The  spinor derivatives $D^k_\a$ and $\bar D_{\da k}$  in these coordinates are
\be
D^k_\a=\pa^k_\a+2i\bt^{\da k}\pada,\q\bar D_{\da k}=\bar\pa_{\da k}.
\ee

Using harmonic projections of the Grassmann coordinates $\th^{\pm\a}=u^\pm_k
\th^{\a k}$ and $\bt^{\pm\da}=u^\pm_k\bt^{\da k}$ one can define the analytic 
coordinates $(x_A, \th^\pm, \bt^\pm)$. The corresponding representation of spinor 
derivatives and supersymmetry generators can be found in \cite{ILZ,FILSZ}. The 
bosonic part of the harmonic superspaces is $R^4\times S^2$, and the left-right 
Grassmann dimensions of general, chiral or analytic superspaces are (4,4), (4,0) 
or (2,2), respectively. We shall use the notation S$(4,2|4,4)$  for the 
supercommutative algebra  and  S$_\star(4,2|4,4)$ for the non-anticommutative 
algebra of general harmonic superfields, respectively. 

It is convenient to use the following differential representation of the SUSY(1,1) 
generators  on S$(4,2|4,4)$:
\bea
&&  T_l^k=-\th^\a_l\pa^k_\a+\sfrac12\d^k_l\th^\a_j\pa^j_\a+\bt^{\da k}
\bar\pa_{\da l}-\sfrac12\d^k_l\bt^{\da j}\bar\pa_{\da j}-u^\pm_l\pa^{\mp k}
+\sfrac12\d^k_l u^\pm_j\pa^{\mp j},\nn\\
&& L_\a^\b=L^\b_\a(y)+\th^\b_k\pa_\a^k-\sfrac12\d^\b_\a\th^\g_k\pa_\g^k,\q 
R^\db_\da=R^\db_\da(y)+\bt^{\db k}\bar\pa_{\da k}-\sfrac12\d^\db_\da\bt^{\dg k}
\bar\pa_{\dg k},\nn\\
&& O=\th^\a_k\pa_\a^k-\bt^{\da k}\bar\pa_{\da k},\q Q_\a^k=\pa^k_\a,\q 
\bar Q_{\da k}=\bar\pa_{\da k}-2i\th^\a_k\pada,\q P_m=\pa_m\lb{dif11}
\eea
where $L^\b_\a(y)$ and $R^\db_\da(y)$ are defined above \p{difgen}, and partial 
harmonic derivatives act as follows $\pa^{\mp l}u^\pm_k=\d^l_k$. For our purposes, 
it is convenient to consider  the following combinations of SUSY(1,1) generators 
and corresponding parameters:
\bea
&&g=P_c+R_\r+Q_\eps,\q G=T_u+L_\l+\bar Q_\bep+aO,\\
&&P_c=c_mP_m,\q T_u=u^k_lT^l_k,\q L_\l=\l^\a_\b L^\b_\a,\q R_\r=\r^\da_\db 
R^\db_\da,\q Q_\eps=\eps^\a_kQ^k_\a,\q \bar Q_\bep=\bep^{\da k}\bar Q_{\da k}.\nn
\eea

The  $N{=}(1,1)$ twist operator $\cF=\exp{(\cP)}$ contains the nilpotent operator
\be
\cP=-\sfrac12C^\ab_{kl}Q^k_\a\otimes Q^l_\b,\q\cP^5=0\lb{PQdef}
\ee
where $C^\ab_{kl}$ are some constants.  The non-anticommutative product in the
corresponding deformed algebra S$_\star(4,2|4,4)$ can be defined by  equivalent
formulas
\bea
&&A\star B=A\exp(P)B=\mu\circ\exp{(\cP)}A\otimes B=\mu_\star\circ A\otimes B
\eea
where $\mu$ and $\mu_\star$ are  product maps for S$(4,2|4,4)$ and 
S$_\star(4,2|4,4)$ and $P$ is the basic operator from \cite{ILZ,FS}
\be
A PB=-\sfrac12(-1)^{p(A)}C^\ab_{kl}Q^k_\a A\, Q^l_\b B=\mu\circ\cP\,A\otimes B.
\ee
The non-anticommutative algebras of the $N{=}(1,1)$ chiral or G-analytic 
superfields can be defined analogously.

By  analogy with Eq.\p{mapdif}, one can construct  the image operator $\hX_{D_1}$ 
on  S$_\star(4,2|4,4)$ for any 1-st order differential operator $D_1=\xi^\M(z)
\pa_\M$  on S$(4,2|4,4)$. The twisted coproduct of SUSY$_t$(1,1) $\D_t(G)=
e^{-\cP} \D(G)e^\cP$ is deformed on  $G=(\bar{Q}_\bep+ T_u+ L_\l+a O)$, in 
particular, 
\bea
&&\D_t(\bar Q_\bep)=\bar Q_\bep\otimes 1+1\otimes \bar Q_\bep+i\bep^{\da k}
C^\ab_{kj}\pada
\otimes Q^j_\b-i\bep^{\da k}C^\ab_{ik}Q^i_\a\otimes  \pa_{\b\da},\nn\\
&&\D_t(T_u)=T_u\otimes 1+1\otimes T_u-\sfrac12u^l_kC^\ab_{lj}Q^k_\a\otimes Q^j_\b
-\sfrac12u^l_kC^\ab_{jl}Q^j_\a\otimes  Q^k_\b,
\nn\\
&&\D_t(L_\l)=L_\l\otimes 1+1\otimes L_\l+\sfrac12\l^\a_\b C^{\b\g}_{kl} Q^k_\a
\otimes Q^l_\g+\sfrac12\l^\a_\b C^{\r\b}_{kl} Q^k_\r\otimes  Q^l_\a,\nn\\
&&\D_t( O)=O\otimes 1+1\otimes O-C^\ab_{kl}Q^k_\a\otimes Q^l_\b,\lb{copr11}
\eea
while $e^{-\cP}\D(g)e^\cP=\D(g)$ for $g=Q_\eps+P_c+R_\r$.

The $\star$-products of arbitrary $N{=}(1,1)$ superfields preserve covariance with 
respect to all deformed transformations of SUSY$_t$(1,1)
\bea
&&\hd_G\star(A\star B)=- \mu_\star\circ \D_t(G)A\otimes B=-G(A\star B).
\eea
The deformed Leibniz rules can be derived directly from these covariant relations.

The twisted supersymmetry acts noncovariantly on the supercommutative product
of superfields, for instance,
\bea
&&\hd_\bep\star(A B)\equiv -\mu\circ \D_t(\bar Q_\bep)A\otimes B\nn\\
&&=-\bar Q_\bep(AB)-i\bep^{\da k}C^\ab_{kj}(-1)^{p(A)}\pada A\, Q^j_\b B+
i\bep^{\da k}C^\ab_{ik}Q^i_\a A\,  \pa_{\b\da}B.
\eea
It is easy to define the $t$-deformed transformations on the products of the 
$N{=}(1,1)$ component fields using the corresponding Grassmann decompositions.

In the special case of the singlet deformation \cite{FILSZ,ILZ2}, the twist 
operator is defined by the parameter $I$ and the SU(2)$\times$SU(2)$_\sL$ 
invariant constant tensor
\be
C^\ab_{kl}=2I\ve^\ab\ve_{kl}~\Rightarrow~\cP_s=-IQ^i_\a\otimes Q_i^\a .
\ee
The $\cP_s$-twist deformation vanishes  for SU(2) and SU(2)$_\sL$ transformations.

The Leibniz rules for differential operators $D=(\pa_m, D^k_\a,\bar D_{\da k},
\pa/\pa u^\pm_k)$ are standard for the general $Q$-deformation \p{PQdef}
\be
D(A\star B)=(DA\star B)+(-1)^{p(D)p(A)}(A\star DB).
\ee

The $\star$-product preserves differential constraints of chirality, antichirality
and Grassmann analyticity \cite{ILZ,FS}. All superfield actions using  
$\star$-products in the non-anticommutative $N=(1,1)$ harmonic superspace 
\cite{FILSZ,ILZ2} are invariant with respect to the quantum group SUSY$_t$(1,1), 
and this invariance is a natural basic principle of these deformed theories. Free 
quadratic parts of these actions possess also the undeformed $N=(1,1)$ 
supersymmetry. The simple examples of the superfield-density terms 
for the analytic hypermultiplet $q^+,~ \tilde q^+$ and the U(1) gauge potential 
$V^{++}$ in the deformed theory are
\be
\tilde q^+\star(\Dp q^+ +[\Vp,q^+]_\star)+\l q^+\star q^+\star\tilde q^+\star
\tilde q^+.
\ee
The $t$-supersymmetry transformations of any term $L^{+4}_\star$ in this density 
are
\bea
&&\hd_G\star L^{+4}_\star=-(\bep^{\da k}\bar Q_{\da k}+u^l_kT^k_l+l^\a_\b 
L^\b_a+aO)L^{+4}_\star,
\eea
so the analytic-superspace integrals of these variations vanish. We hope that the  
manifest SUSY$_t$(1,1) covariance   could help to prove the nonrenormalization 
theorems in $t$-deformed harmonic-superfield theories by analogy with the 
corresponding undeformed theories.

\section{Conclusions}

We analyzed the twist deformations of the Euclidean $N{=}(\sfrac12,\sfrac12)$ and
$N{=}(1,1)$ supersymmetries. By  analogy with the formalism of the deformed 
Minkowski space \cite{We}, we construct explicitly the map between differential 
operators on ordinary and deformed superspaces \p{mapdif}. This map connects the 
standard representation for the supersymmetry generators with the corresponding 
operator representation on the deformed superspace. It is shown  that the 
noncommutative $\star$-products of primary superfields transform covariantly 
in these $t$-deformed supersymmetries. This covariance is a basic principle of 
the superfield formalism of the deformed theories. The Grassmann-coordinate 
decompositions  of the $\star$-product superfields define the deformed tensor
calculus for the components of primary superfields. The ordinary supercommutative 
products of primary superfields or component fields are not covariant with 
respect to the deformed supersymmetries.

Any polynomial terms of the  superfield actions in the non-anticommutative 
$N{=}(\sfrac12,\sfrac12)$ \cite{Se} and  $N{=}(1,1)$ \cite{ILZ,FS} superspaces 
are manifestly invariant with respect to
the corresponding $t$-deformed supersymmetries. The bilinear free parts of these 
actions are also invariant under the standard supersymmetry transformations.
The  deformation constants of the non-anticommutative superfield theories 
break some undeformed (super)symmetries, however, these  parameters can be 
treated as
`coupling constants' compatible with the deformed supersymmetries.  We hope that 
$t$-supersymmetries would help 
to analyze nonrenormalization theorems using the superfield effective actions in 
these theories.

I am grateful to P.P. Kulish for stimulating discussions of  quantum-group 
symmetries in the noncommutative field theory. This work was partially supported 
by  DFG grant 436 RUS 113/669-2 , by  RFBR grants 03-02-17440 and 04-02-04002, 
by  NATO grant PST.GLG.980302 and by  grants of the Heisenberg-Landau and 
Votruba-Blokhintsev programs.

\end{document}